\documentclass[12pt]{article}
\usepackage{axodraw4j}
\usepackage{color}
\usepackage{epsfig}
\usepackage{latexsym}
\usepackage{amsmath}
\usepackage{amssymb}
\usepackage{amsfonts}
\usepackage{pstricks}
\def\e{{\rm e}}

\newcommand{\be}{\begin{equation}}
\newcommand{\ee}{\end{equation}}
\newcommand{\bea}{\begin{eqnarray}}
\newcommand{\eea}{\end{eqnarray}}

\newcommand{\Gm}{\Gamma}

\newcommand{\ep}{\epsilon}

\newcommand{\dd}{\mbox{d}}

\newcommand{\nn}{\nonumber}

\renewcommand{\Im}{\operatorname{Im}}

\begin{document}
\parindent=1.5pc

\begin{titlepage}

\bigskip
\begin{center}
{{\large\bf
Analytic Results
for Massless Three-Loop Form Factors
} \\
\vglue 5pt \vglue 1.0cm
{\large   R.N. Lee}\footnote{E-mail: R.N.Lee@inp.nsk.su}\\
\baselineskip=14pt \vspace{2mm} {\normalsize Budker Institute of Nuclear Physics and
 Novosibirsk State University, \\630090, Novosibirsk, Russia
}\\
\baselineskip=14pt \vspace{2mm} \baselineskip=14pt \vspace{2mm}
{\large  A.V. Smirnov}\footnote{E-mail: asmirnov80@gmail.com}\\
\baselineskip=14pt \vspace{2mm} {\normalsize Scientific Research Computing Center,
Moscow State University, \\ 119992 Moscow, Russia
   }\\
\baselineskip=14pt \vspace{2mm} \baselineskip=14pt \vspace{2mm}
{\large   V.A. Smirnov}\footnote{E-mail: smirnov@theory.sinp.msu.ru}\\
\baselineskip=14pt \vspace{2mm} {\normalsize
Skobeltsyn Institute of Nuclear Physics of Moscow State University, \\
119992 Moscow, Russia
}\\
\baselineskip=14pt \vspace{2mm} \vglue 0.8cm {Abstract}}
\end{center}
\vglue 0.3cm {\rightskip=3pc
 \leftskip=3pc
\noindent We  evaluate, exactly in $d$, the master integrals contributing to massless
three-loop QCD form factors. The calculation is based on a combination of a method recently
suggested by one of the authors (R.L.) with other techniques: sector decomposition
implemented in {\tt FIESTA}, the method of Mellin--Barnes representation, and the PSLQ
algorithm. Using our results for the master integrals we obtain analytical expressions for two
missing constants in the $\ep$-expansion of the two most complicated  master integrals and
present the form factors in a completely analytic form. \vglue 0.8cm}
\end{titlepage}

\section{Introduction}

Recently the evaluation of the QCD form factors at the three-loop level has attracted much
attention. The form factors constitute important building blocks for a number of physical
applications. Among them are the two-jet cross section in $e^+e^-$ collisions, the Higgs-boson
production in the gluon fusion and  the lepton  pair  production  in  proton collisions via the
Drell-Yan mechanism.
The three-loop corrections to the form factors
of the photon-quark and the effective gluon-Higgs boson vertex
appear after integrating out the heavy top-quark loops.
Let $\Gamma^{\mu}_q$ and $\Gamma^{\mu\nu}_g$ be the corresponding vertex functions.
Then the form factors are defined by
\begin{eqnarray}
  F_q(q^2) &=& -\frac{1}{4(1-\ep)q^2}
  \mbox{Tr}\left( q_2\!\!\!\!\!/\,\,\, \Gamma^\mu_q q_1\!\!\!\!\!/\,\,\,
    \gamma_\mu\right)
  \,, \;\;\; \label{Fq}
  \\
  F_g(q^2) &=&
  \frac{\left(q_1\cdot q_2\,\,
      g_{\mu\nu}-q_{1,\mu}\,q_{2,\nu}-q_{1,\nu}\,q_{2,\mu}\right)}
  {2(1-\ep)}
  \Gamma^{\mu\nu}_g \label{Fg}
 \,,
\end{eqnarray}
where $q_1$ and $-q_2$ are the momenta of the incoming and outgoing particles (quarks, for
the case of $F_q$, and gluons, for the case of $F_g$), and  $q=q_1+q_2$ is the momentum
transfer. Here and below, if not stated otherwise, we put $d=4-2\ep$.
Within perturbative expansion, the form factors take the form
\begin{eqnarray}
  F_x &=& 1 + \sum_n \left(\frac{\alpha_s}{4\pi}\right)^n
  \left(\frac{\mu^2}{Q^2}\right)^{n\ep} F_x^{(n)}
  \;, \label{Fx}
\end{eqnarray}
where $Q^2=-q^2$, and $x$ is either $q$(quark) or  $g$(gluon). One deals with the three-loop
order and splits $F_q^{(3)}$ into the singlet, fermionic and remaining gluonic part
\begin{eqnarray}
  F_q^{(3)} &=&  F_q^{(3),g} + F_q^{(3),n_f} +\sum_{q^\prime} Q_{q^\prime} F_q^{(3),sing}
  \,,
  \label{Fq3}
\end{eqnarray}
where $n_f$ stands for the number of active quarks, and $Q_q$ is the charge of the quark $q$.
The pole parts of $F_q^{(3),g}$ and $F_g^{(3)}$ in $\ep$  were presented in Eqs.~(3.7) of
Ref.~\cite{Moch:2005id} and Eqs.~(9) of Ref.~\cite{Moch:2005tm}, respectively.
The finite parts  $F_q^{(3),g}$, $F_q^{(3),sing}$ and $F_g^{(3)}$
were presented in Ref.~\cite{3lff3}.


The integration-by-part reduction reduces the problem to the calculation
of a small number of master integrals. All the master integrals apart from three most
complicated master integrals contributing to the three-loop massless form factors have been
evaluated in \cite{3lff1,3lff2}. In fact, the word {\em evaluated} means here the evaluation up to
the order of $\ep$ which appears in the finite part of the form factors. Mathematically, this
means the evaluation up to transcendentality weight six. About one year ago, one of the three
most complicated master integrals (called $A_{9,1}$ in \cite{3lff1,3lff2,3lff3,3lff4}) and the pole
parts of $A_{9,4}$ and $A_{9,2}$ (shown in Figs.~1 and~2 in the next section) were
evaluated analytically, while the $\ep^0$ parts of
$A_{9,4}$ and $A_{9,2}$ were evaluated numerically --- see \cite{3lff3,3lff4}. Therefore, only
the two (apparently, most complicated) pieces of the whole family of three-loop massless form
factor master integrals are missing at the moment.
Mathematically and aesthetically, it is desirable to obtain completely analytic results, and
this is the problem we are going to solve in the present paper.

Recently, in Ref. \cite{Lee:2009dh} a method of multiloop calculations  based on
the use of dimensional recurrence relations (DRR) \cite{Tarasov1996} and analytic properties
of Feynman integrals as functions of the parameter of dimensional regularization, $d$,
has been suggested. In the present paper we apply
this method to evaluate, exactly in $d$, the master integrals contributing to massless
three-loop QCD form factors. Using the derived expressions, we obtain analytic
results for the missing two constants and thereby arrive at analytic expressions for the
form factors.

The key point of the approach of  Ref. \cite{Lee:2009dh} is the analysis of the analytic
properties of a given integral in a basic stripe of the complex plane $d$. The proper choice of
the master integral, the basic stripe, and the summation factor can essentially simplify the
analysis reducing the number of (or totally fixing) the constants parametrizing the
homogeneous solution of DRR. The freedom of this choice, being an advantage, is also the
only heuristic part of the method. For the case of massive tadpoles, this choice is relatively
simple due to the possibility to get rid of the infrared and ultraviolet singularities by performing
an analysis in an infrared-safe region $d\in(d_0,d_0+2)$ and raising, if necessary, the powers
of the massive denominators (see, e.g., example~2 in Ref. \cite{Lee:2009dh}). For the case of
massless on-shell vertex integrals this recipe does not necessarily work because raising the
powers of the massless denominators also makes worse the infrared and collinear behavior of the
integral. Thus, in this case, one should rely on an analysis of the corresponding parametric
representation. A manual analysis of the parametric representation for the purpose of revealing
the position and the order of the poles can still be a very complicated problem for the cases
considered in this paper. Fortunately, the current version of the code \texttt{FIESTA}
based on sector decompositions provides the possibility to solve this problem automatically.
So, in order to apply the method of Ref. \cite{Lee:2009dh} to
the calculation of a given master integral, we apply a complete set of various techniques:

({\em i}) a reduction to master integrals by two alternative ways: by a code based on
\cite{Lee:2008tj} and the code called {\tt FIRE} \cite{FIRE} to obtain DRR,

({\em ii}) a sector
decomposition \cite{BH,BognerWeinzierl,FIESTA} implemented in the code {\tt FIESTA}
\cite{FIESTA,FIESTA2} to determine the position and the order of the poles in the basic
stripe,

({\em iii}) the method of Mellin--Barnes representation \cite{MB1,MB2,books2} to fix the remaining
constants parametrizing the homogeneous solution (if any),

({\em iv}) PSLQ \cite{PSLQ} to guess the analytical expression for both the constants
parametrizing the homogeneous solution and for the $\ep$-expansion of the master integral
around $d=4$.

As a result, we obtain representations for all master integrals in arbitrary $d$. The
representations have the form of convergent series which allow, in particular, a fast
high-precision calculation of the $\ep$-expansion around $d=4$.

The paper is organized as follows. In the next section we present an example of the
calculation for the integral $A_{7,2}$ and give exact results for this integral and for the lower
master integral $A_{6,3}$. We also present analytical expressions for the $\ep$-expansion of
the integrals $A_{9,2}$ and $A_{9,4}$.

In the conclusion, starting from results for the form factors of Ref.~\cite{3lff3} and substituting
the two constants by our analytic values, we present completely analytic expressions for the
finite parts of the form factors.

\begin{figure}
\begin{center}
\fcolorbox{white}{white}{
  \begin{picture}(290,362) (23,-11)
    \SetWidth{1.0}
    \SetColor{Black}
    \Arc[clock](60,314)(24,-180,-360)
    \Arc[clock](60,296)(30,143.13,36.87)
    \Arc(60,332)(30,-143.13,-36.87)
    \Arc(60,314)(24,-180,0)
    \Line(24,314)(36,314)
    \Line(84,314)(96,314)
    \Line(120,314)(132,314)
    \Line(132,314)(168,350)
    \Line(132,314)(168,278)
    \Line(156,338)(156,290)
    \Arc[clock](128.5,314)(36.5,41.112,-41.112)
    \Arc(183.5,314)(36.5,138.888,221.112)
    \Line(192,314)(204,314)
    \Line(204,314)(240,350)
    \Line(204,314)(240,278)
    \Arc[clock](200.5,314)(36.5,41.112,-41.112)
    \Line(228,338)(228,290)
    \Arc(225,311)(21.213,171.87,278.13)
    \Line(60,122)(96,158)
    \Line(60,122)(96,86)
    \Line(72,110)(84,146)
    \Line(84,98)(72,134)
    \Line(48,122)(60,122)
    \Arc[clock](81,137)(9.487,-161.565,-288.435)
    \Line(48,26)(60,26)
    \Line(60,26)(96,62)
    \Line(60,26)(96,-10)
    \Line(72,26)(84,26)
    \Line(72,26)(84,50)
    \Line(84,26)(72,38)
    \Line(72,26)(72,14)
    \Line(84,26)(84,2)
    \Vertex(36,314){1.414}
    \Vertex(84,314){1.414}
    \Vertex(132,314){1.414}
    \Vertex(156,338){1.414}
    \Vertex(156,290){1.414}
    \Vertex(204,314){1.414}
    \Vertex(228,290){1.414}
    \Vertex(228,338){1.414}
    \Vertex(60,122){1.414}
    \Vertex(72,134){1.414}
    \Vertex(84,146){1.414}
    \Vertex(72,110){1.414}
    \Vertex(84,98){1.414}
    \Vertex(72,38){1.414}
    \Vertex(84,50){1.414}
    \Vertex(72,26){1.414}
    \Vertex(84,26){1.414}
    \Vertex(72,14){1.414}
    \Vertex(60,26){1.414}
    \Vertex(84,2){1.414}
    \Text(24,278)[lb]{\Large{\Black{$A_4$}}}
    \Text(120,278)[lb]{\Large{\Black{$A_{5,1}$}}}
    \Text(192,278)[lb]{\Large{\Black{$A_{5,2}$}}}
    \Text(48,86)[lb]{\Large{\Black{$A_{7,2}$}}}
    \Text(48,-10)[lb]{\Large{\Black{$A_{9,4}$}}}
    \Line(276,314)(312,350)
    \Line(276,314)(312,278)
    \Line(264,314)(276,314)
    \Arc(309,326)(15,126.87,233.13)
    \Arc[clock](291,326)(15,53.13,-53.13)
    \Arc(309,302)(15,126.87,233.13)
    \Arc[clock](291,302)(15,53.13,-53.13)
    \Vertex(276,314){1.414}
    \Vertex(300,338){1.414}
    \Vertex(300,314){1.414}
    \Vertex(300,290){1.414}
    \Text(264,278)[lb]{\Large{\Black{$A_{6,1}$}}}
    \Line(48,218)(60,218)
    \Line(60,218)(96,254)
    \Line(60,218)(96,182)
    \Line(84,242)(84,194)
    \Line(72,218)(84,194)
    \Line(72,218)(84,242)
    \Line(60,218)(72,218)
    \Line(120,218)(132,218)
    \Line(132,218)(168,254)
    \Line(132,218)(168,182)
    \Line(156,242)(156,194)
    \Line(144,230)(156,194)
    \Arc[clock](153,233)(9.487,-161.565,-288.435)
    \Vertex(60,218){1.414}
    \Vertex(72,218){1.414}
    \Vertex(84,242){1.414}
    \Vertex(84,194){1.414}
    \Vertex(132,218){1.414}
    \Vertex(144,230){1.414}
    \Vertex(156,242){1.414}
    \Vertex(156,194){1.414}
    \Text(48,182)[lb]{\Large{\Black{$A_{6,2}$}}}
    \Text(120,182)[lb]{\Large{\Black{$A_{6,3}$}}}
    \Line(192,218)(204,218)
    \Line(204,218)(240,182)
    \Line(204,218)(240,254)
    \Line(228,194)(216,230)
    \Line(216,206)(228,242)
    \Line(216,230)(216,206)
    \Vertex(204,218){1.414}
    \Vertex(216,230){1.414}
    \Vertex(228,242){1.414}
    \Vertex(216,206){1.414}
    \Vertex(228,194){1.414}
    \Text(192,182)[lb]{\Large{\Black{$A_{7,4}$}}}
    \Line(132,122)(168,158)
    \Line(120,122)(132,122)
    \Line(132,122)(168,86)
    \Line(156,146)(156,98)
    \Line(156,98)(144,134)
    \Line(144,110)(156,146)
    \Vertex(132,122){1.414}
    \Vertex(144,134){1.414}
    \Vertex(156,146){1.414}
    \Vertex(144,110){1.414}
    \Vertex(156,98){1.414}
    \Text(120,86)[lb]{\Large{\Black{$A_{7,5}$}}}
  \end{picture}
}
\end{center} 
\caption[]{Master integrals for $A_{9,4}$.}
\end{figure}
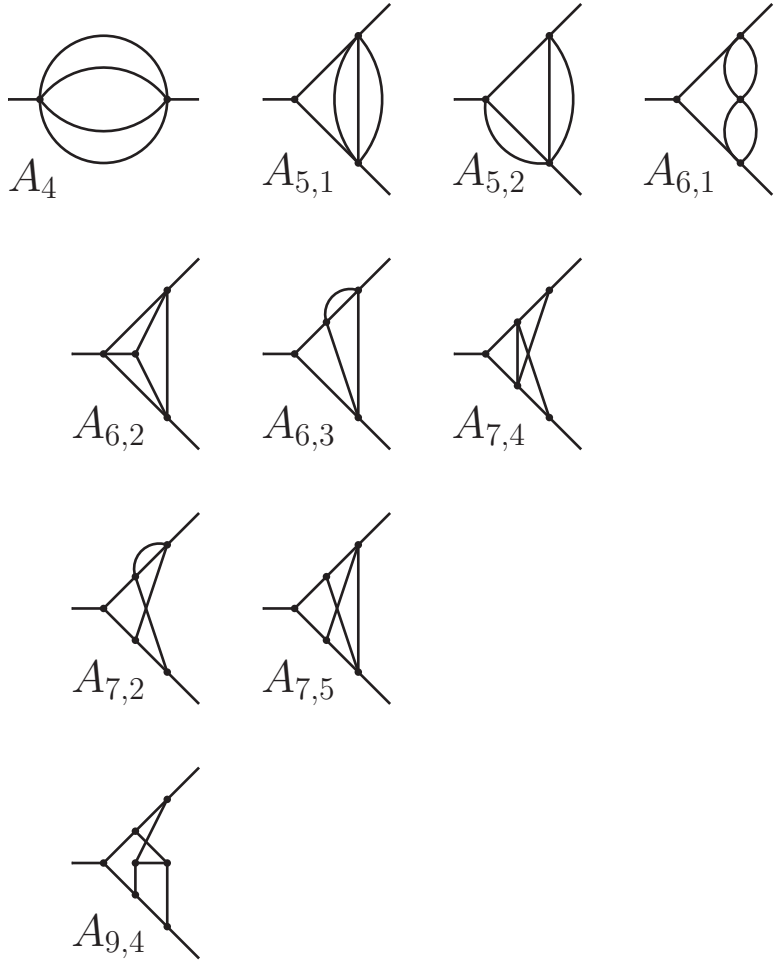 
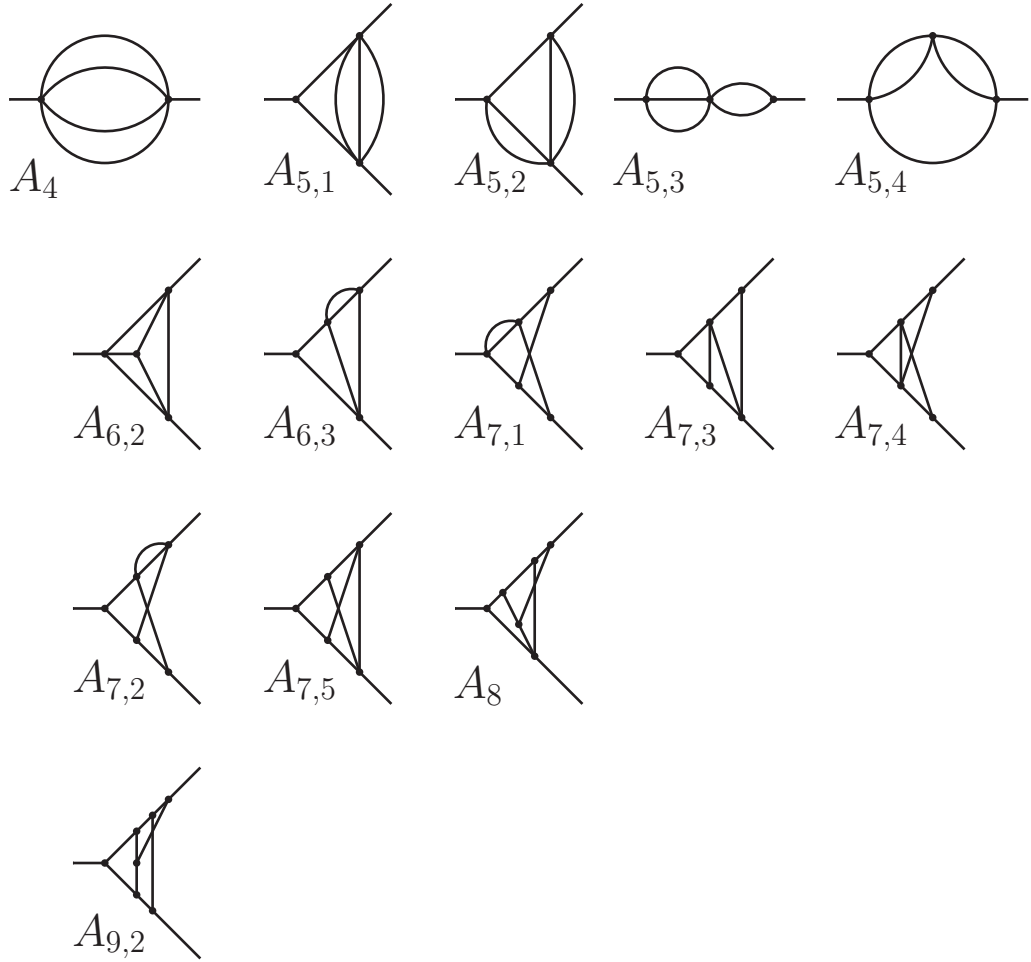
\begin{figure}
\begin{center}
\fcolorbox{white}{white}{
  \begin{picture}(386,362) (23,-11)
    \SetWidth{1.0}
    \SetColor{Black}
    \Arc[clock](60,314)(24,-180,-360)
    \Arc[clock](60,296)(30,143.13,36.87)
    \Arc(60,332)(30,-143.13,-36.87)
    \Arc(60,314)(24,-180,0)
    \Line(24,314)(36,314)
    \Line(84,314)(96,314)
    \Line(120,314)(132,314)
    \Line(132,314)(168,350)
    \Line(132,314)(168,278)
    \Line(156,338)(156,290)
    \Arc[clock](128.5,314)(36.5,41.112,-41.112)
    \Arc(183.5,314)(36.5,138.888,221.112)
    \Line(192,314)(204,314)
    \Line(204,314)(240,350)
    \Line(204,314)(240,278)
    \Arc[clock](200.5,314)(36.5,41.112,-41.112)
    \Line(228,338)(228,290)
    \Line(48,218)(60,218)
    \Line(60,218)(96,254)
    \Line(60,218)(96,182)
    \Line(84,242)(84,194)
    \Line(72,218)(84,194)
    \Line(72,218)(84,242)
    \Line(60,218)(72,218)
    \Line(120,218)(132,218)
    \Line(132,218)(168,254)
    \Line(132,218)(168,182)
    \Line(156,242)(156,194)
    \Line(144,230)(156,194)
    \Arc[clock](153,233)(9.487,-161.565,-288.435)
    \Arc(225,311)(21.213,171.87,278.13)
    \Vertex(36,314){1.414}
    \Vertex(84,314){1.414}
    \Vertex(132,314){1.414}
    \Vertex(156,338){1.414}
    \Vertex(156,290){1.414}
    \Vertex(204,314){1.414}
    \Vertex(228,290){1.414}
    \Vertex(228,338){1.414}
    \Vertex(60,218){1.414}
    \Vertex(72,218){1.414}
    \Vertex(84,242){1.414}
    \Vertex(84,194){1.414}
    \Vertex(132,218){1.414}
    \Vertex(144,230){1.414}
    \Vertex(156,242){1.414}
    \Vertex(156,194){1.414}
    \Text(24,278)[lb]{\Large{\Black{$A_4$}}}
    \Text(120,278)[lb]{\Large{\Black{$A_{5,1}$}}}
    \Text(192,278)[lb]{\Large{\Black{$A_{5,2}$}}}
    \Text(48,182)[lb]{\Large{\Black{$A_{6,2}$}}}
    \Text(120,182)[lb]{\Large{\Black{$A_{6,3}$}}}
    \Line(252,314)(264,314)
    \Arc[clock](276,314)(12,-180,-360)
    \Arc[clock](276,314)(12,-0,-180)
    \Arc[clock](300,305)(15,143.13,36.87)
    \Arc(300,323)(15,-143.13,-36.87)
    \Line(264,314)(288,314)
    \Line(312,314)(324,314)
    \Line(336,314)(348,314)
    \Line(396,314)(408,314)
    \Arc(372,314)(24,180,540)
    \Arc(400,342)(28.284,-171.87,-98.13)
    \Arc(344,342)(28.284,-81.87,-8.13)
    \Vertex(264,314){1.414}
    \Vertex(288,314){1.414}
    \Vertex(312,314){1.414}
    \Vertex(348,314){1.414}
    \Vertex(396,314){1.414}
    \Vertex(372,338){1.414}
    \Text(252,278)[lb]{\Large{\Black{$A_{5,3}$}}}
    \Text(336,278)[lb]{\Large{\Black{$A_{5,4}$}}}
    \Line(192,218)(204,218)
    \Text(192,182)[lb]{\Large{\Black{$A_{7,1}$}}}
    \Line(204,218)(240,182)
    \Line(204,218)(240,254)
    \Arc[clock](213,221)(9.487,-161.565,-288.435)
    \Line(228,194)(216,230)
    \Line(216,206)(228,242)
    \Vertex(204,218){1.414}
    \Vertex(216,230){1.414}
    \Vertex(228,242){1.414}
    \Vertex(216,206){1.414}
    \Vertex(228,194){1.414}
    \Line(336,218)(348,218)
    \Line(348,218)(384,182)
    \Line(348,218)(384,254)
    \Line(372,194)(360,230)
    \Line(360,206)(372,242)
    \Line(360,230)(360,206)
    \Vertex(348,218){1.414}
    \Vertex(360,230){1.414}
    \Vertex(372,242){1.414}
    \Vertex(360,206){1.414}
    \Vertex(372,194){1.414}
    \Text(336,182)[lb]{\Large{\Black{$A_{7,4}$}}}
    \Line(264,218)(276,218)
    \Text(264,182)[lb]{\Large{\Black{$A_{7,3}$}}}
    \Line(276,218)(312,182)
    \Line(276,218)(312,254)
    \Line(288,230)(288,206)
    \Line(300,194)(288,230)
    \Line(300,242)(300,194)
    \Vertex(276,218){1.414}
    \Vertex(288,230){1.414}
    \Vertex(288,206){1.414}
    \Vertex(300,194){1.414}
    \Vertex(300,242){1.414}
    \Line(60,122)(96,158)
    \Line(60,122)(96,86)
    \Line(72,110)(84,146)
    \Line(84,98)(72,134)
    \Line(48,122)(60,122)
    \Arc[clock](81,137)(9.487,-161.565,-288.435)
    \Vertex(60,122){1.414}
    \Vertex(72,134){1.414}
    \Vertex(84,146){1.414}
    \Vertex(72,110){1.414}
    \Vertex(84,98){1.414}
    \Text(48,86)[lb]{\Large{\Black{$A_{7,2}$}}}
    \Line(132,122)(168,158)
    \Line(120,122)(132,122)
    \Line(132,122)(168,86)
    \Line(156,146)(156,98)
    \Line(156,98)(144,134)
    \Line(144,110)(156,146)
    \Vertex(132,122){1.414}
    \Vertex(144,134){1.414}
    \Vertex(156,146){1.414}
    \Vertex(144,110){1.414}
    \Vertex(156,98){1.414}
    \Text(120,86)[lb]{\Large{\Black{$A_{7,5}$}}}
    \Line(192,122)(204,122)
    \Line(204,122)(240,158)
    \Line(204,122)(240,86)
    \Line(210,128)(222,104)
    \Line(222,104)(222,140)
    \Line(216,116)(228,146)
    \Vertex(204,122){1.414}
    \Vertex(210,128){1.414}
    \Vertex(216,116){1.414}
    \Vertex(222,104){1.414}
    \Vertex(222,140){1.414}
    \Vertex(228,146){1.414}
    \Text(192,86)[lb]{\Large{\Black{$A_{8}$}}}
    \Line(48,26)(60,26)
    \Line(60,26)(96,62)
    \Line(60,26)(96,-10)
    \Vertex(72,38){1.414}
    \Vertex(84,50){1.414}
    \Vertex(72,14){1.414}
    \Vertex(60,26){1.414}
    \Text(48,-10)[lb]{\Large{\Black{$A_{9,2}$}}}
    \Line(72,38)(72,14)
    \Line(72,26)(84,50)
    \Line(78,44)(78,8)
    \Vertex(72,26){1.414}
    \Vertex(78,8){1.414}
    \Vertex(78,44){1.414}
  \end{picture}
}
\end{center} 
\caption[]{Master integrals for $A_{9,2}$.}
\end{figure} 

\section{Master Integrals for Massless \\ Three-Loop Form Factors}

Master integrals naturally form a partially ordered set. One master integral is said to be
\textit{lower} than the other master integral if the Feynman graph for the former can be obtained
by contracting some
internal lines from the Feynman graph of the latter. This ordering enables
us to introduce the notion of \textit{complexity level} of a given master integral
which is the maximal number of nested lower master integrals.
Owing to this definition, the master integrals with zero complexity level have no lower master
integrals. The DRR for such integral is obviously homogeneous and its explicit solution is
expressed in terms of $\Gamma$-functions. Moreover, it turns out that for three-loop on-shell
massless vertex master integrals any integral expressed in terms of $\Gamma$-functions has
zero complexity level. We expect this situation to be general.

Our primary goal is the calculation of the most complicated integrals, $A_{9,2}$ and $A_{9,4}$
which are the last integrals in Figs.~1 and~2.
However, in order to be able to apply the method of Ref. \cite{Lee:2009dh} we have to know
all lower master integrals which are shown in the same figures.
Four rows of diagrams in each figure correspond to complexity levels 0, 1, 2 and 3.
Therefore, we start our calculation from the complexity level~$1$,
then pass to the complexity level $2$ and, finally, calculate the two master integrals of
complexity level $3$. Let us demonstrate an intermediate step of this procedure using the
example of the integral $A_{7,2}$. We directly follow the path of Ref. \cite{Lee:2009dh}:

\begin{enumerate}
\item
There are four lower master integrals, $A_{4}$, $A_{5,1}$, $A_{5,2}$, and $A_{6,3}$. Three
of them are expressed in terms of $\Gamma$-functions, while the last one, $A_{6,3}$, can
be obtained using the same method with the final result conveniently represented as
\begin{eqnarray}
A_{6,3}(d)&=&A_{6,3}^{1,1}(d)\sum_{k=0}^{\infty}A_{6,3}^{1,2}(d+2k)+A_{6,3}^{2}(d)\,,
\nonumber\\
A_{6,3}^{1,1}(d)&=&- \sin (\pi  d)A_{6,3}^2(d)=\frac{\pi ^4 2^{11-3 d}\csc \left(\frac{3 \pi
d}{2}\right)\csc \left(\frac{\pi  d}{2}\right)} {(3 d-10)
\Gamma \left(d-\frac{5}{2}\right) \Gamma \left(\frac{d-1}{2}\right)}\,,
\nonumber\\
A_{6,3}^{1,2}(d)&=&
\frac{(7 d-18)\sin \left(\frac{\pi  d}{2}\right)\Gamma\left(\frac{d}{2}-1\right)^3}
{3\pi^2(d-3)\Gamma \left(\frac{3 d}{2}-3\right)}\,.
\end{eqnarray}
\item
Here and in what follows, we omit, for brevity, a power-like dependence of the master integrals
on $q^2+i0$ which can easily be restored by power counting.

Using the \texttt{FIESTA} program we determine the position and the order of the poles in
the basic stripe which we choose as $S=\left\{d|\quad\operatorname{Re}d\in \left(
4,6\right]\right\}$.
The syntax for this analysis is {\tt
SDAnalyze[{U,F,h},degrees,order,dmin,dmax]},
where {\tt U} and {\tt F} are the basic functions in the parametric integral
corresponding to the given Feynman integral, {\tt h} is the
number of loops, {\tt degrees} are the indices, {\tt order} is the required order in
$\ep$ and {\tt dmin} and {\tt dmax} are values of the real part of $d$ that determine
the basic stripe. The output lists the values of $d$ where the given Feynman integral
can have poles.
This feature appeared in the second version of FIESTA, but was not
documented in \cite{FIESTA2} because testing was still in progress.
So, after applying this procedure to $A_{7,2}$ we see\footnote{In fact, the overall factor
$\Gm(a-hd/2)$ where $a$ is the sum of the indices is not taken into account by
FIESTA but this can easily be done
because the corresponding poles are explicit. Let us emphasize that FIESTA can
report also on some fictitious poles. This can happen when contributions of individual
sectors do have some additional poles which cancel in the sum. However, FIESTA itself can
be used further to check whether the poles are indeed present or not.}
that the integral has simple poles at $d=14/3,5,16/3,6$.

\item
The dimensional recurrence reads\footnote{We use the integration measure  $\dd^dk/(i\pi^{d/2})$ per loop.}%
\begin{eqnarray}
A_{7,2}(d+2)&=&
c_{7,2}(d)A_{7,2}(d)
\\
&&+c_{6,3}(d)A_{6,3}(d)
+c_{5,2}(d)A_{5,2}(d)
+c_{5,1}(d)A_{5,1}(d)
+c_{4}(d)A_{4}(d)\nonumber
\end{eqnarray}
where $c_{n}$ are some rational functions of $d$ presented in the Appendix.

\item
Using the explicit form of the coefficient $c_{7,2}(d)$, we choose the summing factor as
\begin{equation}
\Sigma\left(d\right)  =
\frac{(d-3) \cos \left(\frac{\pi  d}{2}\right) \cos \left(\frac{\pi }{6}-\frac{\pi  d}{2}\right) \cos \left(\frac{\pi
   d}{2}+\frac{\pi }{6}\right)
   \Gamma \left(\frac{5 d}{2}-9\right)}{\Gamma \left(\frac{d}{2}-2\right)^2}\,.
   \label{eq:sf72}
\end{equation}
Passing to the function $\tilde{A}_{7,2}(d)=\Sigma(d)A_{7,2}(d)$, we obtain the following
equation
\begin{eqnarray}
\tilde{A}_{7,2}(d+2)&=&
\tilde{A}_{7,2}(d)+\tilde{A}_{6,3}(d)
+\tilde{A}_{5,2}(d)
+\tilde{A}_{5,1}(d)
+\tilde{A}_{4}(d)\,,
\end{eqnarray}
where $\tilde{A}_{n}(d)=\Sigma(d+2)c_{n}(d)A_{n}(d)$. The general solution can easily be
constructed using the explicit form of the integrals $A_4$, $A_{5,1}$,  $A_{5,2}$, and
$A_{6,3}$:
\begin{eqnarray}
\tilde{A}_{7,2}(d)&=& \omega(z)+\sum_{l=0}^{\infty}
\left[\tilde{A}_{5,2}(d-2-2l)
+\tilde{A}_{5,1}(d-2-2l)
+\tilde{A}_{6,3}^{2}(d-2-2l)
\right]
\nonumber\\
&&
-\sum_{l=0}^{\infty}\tilde{A}_{6,3}^{1,1}(d+2l)\sum_{k=0}^{\infty}A_{6,3}^{1,2}(d+2l+2k)
-\sum_{l=0}^{\infty}\tilde{A}_{4}(d+2l),
\label{eq:sol72}
\end{eqnarray}
where $z=\exp[i\pi d]$.
\item
The function $\Sigma(d)$ has simple zeros at $d=14/3,5,16/3$, therefore,
$\tilde{A}_{7,2}(d)$ is regular everywhere in $S$ except the point $d=6$, where it has a
simple pole. Besides, from the explicit form of the summing factor and from the parametric
representation of $A_{7,2}(d)$ it is immediately clear that $\tilde{A}_{7,2}(d)$ grows slower
than any positive (negative) power of $|z|$ when $\Im d\to -\infty$ ( $\Im d\to +\infty$). This
fixes $\omega(z)$ up to a function
\begin{equation}
a_1+a_2\cot\left(\frac{\pi}{2}(d-6)\right)
\end{equation}
\item
In order to fix the two remaining constants, we use data obtained from the Mellin--Barnes
representation of $A_{7,2}(d)$ which can easily be obtained from the
general Mellin--Barnes representation for the non-planar on-shell vertex diagram
(see, e.g., Chap.~4 of \cite{books2}):
\bea
A_{7,2}(d)=\frac{1}{(2\pi)^2}\int\int\frac{\Gamma \left(\frac{d}{2}-2\right) \Gamma \left(\frac{d}{2}-1\right)^2
   \Gamma (d-3) \Gamma (-z_1) \Gamma (-z_2) \Gamma(z_2+1)^2
 }{\Gamma (d-2) \Gamma \left(\frac{3 d}{2}-5\right) \Gamma (2 d-7) \Gamma
   (d-z_1-4)}
\nn && \\  && \hspace*{-125mm} \nn
\times  \frac{\Gamma \left(\frac{d}{2}-z_1-2\right)}
{ \Gamma \left(\frac{3 d}{2}-z_1-5\right)}
\Gamma \left(\frac{3 d}{2}-z_2-6\right) \Gamma (z_1+z_2+1)
   \Gamma (d-z_1-z_2-5)
\\  && \hspace*{-125mm}
\times\Gamma \left(\frac{3
   d}{2}-z_1-z_2-6\right) \Gamma \left(-\frac{3
   d}{2}+z_1+z_2+7\right) \dd z_1\, \dd z_2\,.
\label{A72MB}
\eea
Using the codes of Refs.~\cite{MB2}, at $d=6-2\ep$ and $d=5-2\ep$
we straightforwardly obtain
\begin{eqnarray}
A_{7,2}(6-2\ep)&=&-\frac{41}{15552\ep}+O(\ep^0),
\nonumber\\
A_{7,2}(5-2\ep)&=&-\frac{\pi ^{5/2}}{24 \ep }+O(\ep^0).
\label{eq:MB72}
\end{eqnarray}
Using these two values and also taking into account the fact that the singularities of  the
inhomogeneous part should be cancelled, we obtain
\begin{eqnarray}
\omega(z)&=&
\frac{\pi ^3 }{20 \sqrt{5}}\tan \left(\frac{\pi }{10}-\frac{\pi  d}{2}\right)
-\frac{\pi ^3}{36}  \tan \left(\frac{\pi}{6}-\frac{\pi  d}{2}\right)
-\frac{\pi ^3 }{20 \sqrt{5}}\tan \left(\frac{\pi  d}{2}+\frac{\pi }{10}\right)
\nonumber\\
&&
+\frac{\pi ^3}{36} \tan \left(\frac{\pi  d}{2}+\frac{\pi }{6}\right)
+\frac{\pi ^3 }{60} \cot ^3\left(\frac{\pi  d}{2}\right)
+\frac{13\pi^3}{180}  \cot \left(\frac{\pi  d}{2}\right)
\nonumber\\
&&
+\frac{\pi ^3 }{20 \sqrt{5}}\cot \left(\frac{\pi }{5}-\frac{\pi  d}{2}\right)
-\frac{\pi ^3}{20 \sqrt{5}}\cot \left(\frac{\pi  d}{2}+\frac{\pi }{5}\right) \,.
\label{eq:omega72}
\end{eqnarray}
\end{enumerate}
Eqs. (\ref{eq:sol72}), (\ref{eq:omega72}), and (\ref{eq:sf72}) determine our final expression for
$A_{7,2}(d)$.

Two remarks are in order. First, our choice of the summing factor, the basic stripe and the
master integral itself (we could have considered instead, e.g., an integral with some
denominators squared and/or with numerators) may be not the most optimal one. With some
other choice, we might have been able to fix the homogeneous part of the solution entirely
within the method. However, given the number of the integrals to be considered and the
absence of the general recipe for this choice, it was much more convenient to use in such
cases additional data from Mellin--Barnes representations. In fact, for other integrals the number
of the constants to be fixed was not greater than two.

The second remark concerns the double sum in Eq. (\ref{eq:sol72}). Making a shift $k\to k-l$, we
obtain the following triangle sum with the factorized summand:
\begin{equation}
\sum_{l=0}^{\infty}\tilde{A}_{6,3}^{1,1}(d+2l)\sum_{k=l}^{\infty}A_{6,3}^{1,2}(d+2k)\,.
\end{equation}
The factorized form of the summand essentially simplifies the numerical calculation of the sum,
making it possible to organize the calculations without nested do-loops.
Proceeding in the same way for the rest of the integrals, we finally obtain general expressions
for $A_{9,4}$ and $A_{9,2}$. The resulting representations for arbitrary $d$ are too lengthy to
be presented here and can be obtained upon request from the authors. We present here only
analytical results for the expansion of these two integrals around $d=4$ which are most
interesting for physical applications. The expansion for $A_{9,4}$ reads
\begin{eqnarray}
A_{9,4}(4-2\ep) =\e^{-3\gamma_E \ep }\biggl\{ -\frac{1}{9\ep^6}-\frac{8}{9\ep^5}
+\left[ 1+\frac{43 \pi ^2}{108}\right]\frac{1}{\ep^4}
+\left[\frac{109 \zeta (3)}{9}+\frac{14}{9}
+\frac{53 \pi ^2}{27} \right]\frac{1}{\ep^3}
&& \nn \\ &&  \hspace*{-120mm}
+\left[ \frac{608 \zeta (3)}{9}-17-\frac{311 \pi ^2}{108}
-\frac{481 \pi^4}{12960}\right]\frac{1}{\ep^2}
\nn \\ &&  \hspace*{-120mm}
+\left[ -\frac{949 \zeta (3)}{9}-\frac{2975 \pi ^2 \zeta (3)}{108}
+\frac{3463\zeta (5)}{45}+84+\frac{11 \pi ^2}{18}
+\frac{85 \pi ^4}{108}\right]\frac{1}{\ep}
\nn \\ &&  \hspace*{-120mm}
+\left[\frac{434 \zeta (3)}{9}-\frac{299 \pi ^2 \zeta (3)}{3}
-\frac{3115 \zeta(3)^2}{6}+\frac{7868 \zeta (5)}{15}
\right. \nn \\ &&  \hspace*{-113mm}
\left. -339+\frac{77 \pi ^2}{4}
-\frac{2539 \pi^4}{2592}-\frac{247613 \pi ^6}{466560} \right] +O(\ep)\biggr\}
\;,
\label{A94}
\end{eqnarray}
For $A_{9,2}$, we arrive at the following result:
\begin{eqnarray}
A_{9,2}(4-2\ep) = \e^{-3\gamma_E \ep }\biggl\{-\frac{2}{9\ep^6}-\frac{5}{6\ep^5}
+\left[ \frac{20}{9}+\frac{17 \pi ^2}{54}\right]\frac{1}{\ep^4}
&& \nn \\ &&  \hspace*{-85mm}
+\left[\frac{31 \zeta (3)}{3}-\frac{50}{9}
+\frac{181 \pi ^2}{216}  \right]\frac{1}{\ep^3}
\nn \\ &&  \hspace*{-85mm}
+\left[\frac{347 \zeta (3)}{18}+\frac{110}{9}-\frac{17 \pi ^2}{9}
+\frac{119 \pi ^4}{432} \right]\frac{1}{\ep^2}
\nn \\ &&  \hspace*{-85mm}
 +\left[-\frac{514 \zeta (3)}{9}-\frac{341 \pi ^2 \zeta (3)}{36}
+\frac{2507 \zeta(5)}{15}-\frac{170}{9}+\frac{19 \pi ^2}{6}
+\frac{163 \pi ^4}{960} \right]\frac{1}{\ep}
\nn \\ &&  \hspace*{-85mm}
+\left[\frac{1516 \zeta (3)}{9}-\frac{737 \pi ^2 \zeta (3)}{24}-29 \zeta (3)^2
+\frac{2783\zeta (5)}{6}
\right. \nn \\ &&  \hspace*{-75mm}
\left.-\frac{130}{9}+\frac{\pi ^2}{2}
-\frac{943 \pi ^4}{1080}+\frac{195551 \pi ^6}{544320}\right] +O(\ep)\biggr\}
\;.
\label{A92}
\end{eqnarray}

\section{Conclusion}

Eqs.~(\ref{A94}) and (\ref{A92}) enable us to
present completely analytic results for the three-loop corrections to the form factors
defined by Eqs.~(\ref{Fq}--\ref{Fq3}).
Starting from Eqs.~(8--10) of Ref.~\cite{3lff3} and taking into account our analytic
values of the $\ep^0$ terms in (\ref{A94}) and
(\ref{A92}) we obtain the following analytic expressions:
\begin{eqnarray}
  F_q^{(3),g+n_f}\Big|_{\rm fin} =
  C_F^3
  \left[
 2669 \zeta (3)+\frac{61 \pi ^2 \zeta (3)}{6}-\frac{1826 \zeta(3)^2}{3}
 +\frac{4238 \zeta (5)}{5}-\frac{53675}{24}
\right.
&& \nn \\ &&  \hspace*{-133mm}
\left.
   -\frac{13001 \pi
   ^2}{72}+\frac{12743 \pi ^4}{1440}-\frac{9095 \pi ^6}{54432}
  \right]
  +C_A  C_F^2
  \left[
 -\frac{96715 \zeta (3)}{18}+\frac{23 \pi ^2 \zeta (3)}{27}+\frac{1616
   \zeta (3)^2}{3} \right.
 \nn \\ &&  \hspace*{-133mm}
\left.
-\frac{46594 \zeta(5)}{45}+\frac{37684115}{5832}+\frac{664325 \pi^2}{1944}
-\frac{1265467 \pi ^4}{77760}-\frac{18619 \pi ^6}{272160}
  \right]
 \nn \\ &&  \hspace*{-133mm}
  +C_A^2C_F \left[
  \frac{1341553 \zeta (3)}{486}-\frac{355 \pi ^2 \zeta (3)}{27}-\frac{1136
   \zeta (3)^2}{9}+\frac{2932 \zeta(5)}{9}-\frac{52268375}{13122}
\right.
 \nn \\ &&  \hspace*{-133mm}
\left.
   -\frac{383660 \pi^2}{2187}+\frac{152059 \pi ^4}{19440}-\frac{769 \pi ^6}{5103}
  \right]
  +C_F^2n_fT
  \left[
    -\frac{2732173}{1458}
-\frac{45235 \pi ^2}{486}
\right.
\nn \\ &&  \hspace*{-130mm}
\left.
    +\frac{102010 \zeta(3)}{81}
+\frac{8149 \pi ^4}{3888}
-\frac{343 \pi ^2 \zeta (3)}{27}
    +\frac{556 \zeta(5)}{45}
  \right]
\nn \\ &&  \hspace*{-130mm}
  +C_AC_Fn_fT
  \left[
     \frac{17120104}{6561}
+\frac{442961 \pi ^2}{4374}
    -\frac{90148 \zeta(3)}{81}
-\frac{1093 \pi ^4}{486}
+\frac{368 \pi ^2 \zeta (3)}{27}
\right.
\nn \\ &&  \hspace*{-130mm}
\left.
    -\frac{416 \zeta(5)}{3}
  \right]
  +C_Fn_f^2
  T^2
  \left[
    -\frac{2710864}{6561}
-\frac{124 \pi ^2}{9}
    +\frac{12784 \zeta(3)}{243}
-\frac{83 \pi ^4}{1215}
  \right]
\;,
\end{eqnarray}
\begin{eqnarray}
  F_g^{(3)}\Big|_{\rm fin} =
  C_A^3 \left[
-\frac{68590 \zeta (3)}{243}+\frac{77 \pi ^2 \zeta (3)}{108}-\frac{1766
   \zeta (3)^2}{9}+\frac{20911 \zeta
   (5)}{45}+\frac{14474131}{13122}
\right.
&& \nn \\ &&  \hspace*{-141mm}
\left.
   +\frac{307057 \pi ^2}{8748}+\frac{8459
   \pi ^4}{38880}-\frac{22523 \pi ^6}{58320}
  \right]
\nn \\ &&  \hspace*{-141mm}
  +C_A^2n_f T
  \left[
    -\frac{10021313}{6561}
-\frac{37868 \pi ^2}{2187}
    -\frac{1508 \zeta(3)}{27}
+\frac{437 \pi ^4}{1080}
-\frac{439 \pi ^2 \zeta (3)}{27}
    +\frac{6476 \zeta(5)}{45}
  \right]
\nn \\ &&  \hspace*{-141mm}
  +C_FC_A n_f T
  \left[
    -\frac{155629}{243}
-\frac{41 \pi ^2}{9}
    +\frac{23584 \zeta(3)}{81}
-\frac{8 \pi ^4}{45}
+16 \pi ^2 \zeta (3)
+\frac{64 \zeta(5)}{9}
  \right]
\nn \\ &&  \hspace*{-141mm}
  +C_F^2n_f T
  \left[
     \frac{608}{9}
    +\frac{592 \zeta(3)}{3}
    -320 \zeta(5)
  \right]
  + C_F n_f^2 T^2
  \left[
     \frac{42248}{81}
-\frac{32 \pi ^2}{9}
\right.
\nn \\ &&  \hspace*{-141mm}
\left.
    -\frac{2816 \zeta(3)}{9}
-\frac{112 \pi ^4}{135}
  \right]
  +C_An_f^2 T^2
  \left[
     \frac{2958218}{6561}
+\frac{152 \pi ^2}{81}
    +\frac{47296 \zeta(3)}{243}
+\frac{797 \pi ^4}{1215}
  \right]
  \;,
\end{eqnarray}
where $C_F=(N_c^2-1)/(2N_c)$, $C_A=N_c$, $T=1/2$ and
$d^{abc}d^{abc}=(N_c^2-1)(N_c^2-4)/N_c$.

We are confident that this technique can be applied to analytically evaluate master
integrals appearing in various physical problems.

\vspace{0.2 cm}

{\em Acknowledgments.}

This work was supported by the Russian Foundation for Basic Research through grants
08--02--01451 (A.S. and V.S.) and 09--02--00024 (R.L.).

\appendix
\section*{Appendix}
The coefficients of the dimensional recurrence relation for $A_{7,2}$  have the form
\begin{eqnarray}
c_{7,2}&=&-\frac{8 (d-4)^2 (d-3)}{5 (d-2) (d-1) (5 d-18) (5 d-16) (5 d-14) (5 d-12)}\,,
 \nn \\
c_{6,3}&=&
-\frac{(3 d-10) \left(483 d^4-5996 d^3+27684 d^2-56272 d+42432\right)}
{20 (d-2)^2 (d-1) (2 d-5) (5 d-18) (5 d-16) (5 d-14) (5 d-12)}\,,
\nn \\
c_{5,2}&=&
-(d-3)[15 (d-4)(d-2)^2 (d-1) (2 d-5) (3 d-10) (3 d-8) (5 d-18)]^{-1}
\nn \\&&
\times[ (5 d-16)(5 d-14) (5 d-12)]^{-1} \left[12447 d^7-256626 d^6+2261972 d^5
\right.
\nn \\&&
\left.
-11052152 d^4+32339200 d^3-56684032 d^2
+55123200 d-22947840\right]\,,
\nn \\
c_{5,1}&=&-\left[
60 (d-4) (d-2)^2 (d-1) (2 d-5)
   (3 d-10) (5 d-18) (5 d-16)
\right]^{-1}
\nn \\&&
\times \left[(5 d-14) (5 d-12)\right]^{-1}
\left[18909 d^7-384006 d^6+3329804 d^5
\right.
\nn \\&&
\left.
-15982952 d^4+45870976 d^3-78731008 d^2+74846208 d-30412800\right]\,,
\nn \\
c_{4}&=&-
\left[
90 (d-3) (d-2)^2 (d-1) (3 d-10) (3 d-8) (5
   d-16) \right]^{-1}
\nn \\&&
\times \left[(5 d-14) (5 d-12)\right]^{-1}
 \left[38619 d^6-651987 d^5+4575500 d^4
\right.
\nn \\&&
\left.
-17083884 d^3+35791888 d^2-39892032 d+18478080\right]
\,.
\end{eqnarray}


\begin{thebibliography}{99}
\bibitem{Moch:2005id}
  S.~Moch, J.~A.~M.~Vermaseren and A.~Vogt,
  JHEP {\bf 0508} (2005) 049
  [arXiv:hep-ph/0507039].

\bibitem{Moch:2005tm}
  S.~Moch, J.~A.~M.~Vermaseren and A.~Vogt,
  Phys.\ Lett.\  B {\bf 625} (2005) 245
  [arXiv:hep-ph/0508055].

\bibitem{3lff3}
  P.~A.~Baikov, K.~G.~Chetyrkin, A.~V.~Smirnov, V.~A.~Smirnov and M.~Steinhauser,
  Phys.\ Rev.\ Lett.\  {\bf 102} (2009) 212002
  [arXiv:0902.3519 [hep-ph]].

\bibitem{3lff1}
  T.~Gehrmann, G.~Heinrich, T.~Huber and C.~Studerus,
  Phys.\ Lett.\  B {\bf 640}, 252 (2006)
  [hep-ph/0607185].

\bibitem{3lff2}
  G.~Heinrich, T.~Huber and D.~Ma\^{\i}tre,
  Phys.\ Lett.\  B {\bf 662} (2008) 344
  [arXiv:0711.3590 [hep-ph]].

\bibitem{3lff4}
G.~Heinrich, T.~Huber, D.~A.~Kosower and V.~A.~Smirnov,
  Phys.\ Lett.\  B {\bf 678} (2009) 359
  [arXiv:0902.3512 [hep-ph]].

\bibitem{Lee:2009dh}
  R.~N.~Lee,
Nucl.\ Phys.\  B {\bf 830} (2010) 474
  [arXiv:0911.0252 [hep-ph]].

\bibitem{Tarasov1996}
O.~V. Tarasov, Phys.~Rev.~D {\bf 54} (1996) 6479.


\bibitem{Lee:2008tj}
  R.~N.~Lee,
  JHEP {\bf 0807} (2008) 031
  [arXiv:0804.3008 [hep-ph]].

\bibitem{FIRE}
 A.~V.~Smirnov,
  JHEP {\bf 0810}, 107 (2008)
  [arXiv:0807.3243 [hep-ph]].

\bibitem{BH}
T.~Binoth and G.~Heinrich, Nucl. Phys. B, {\bf 585} (2000) 741;
Nucl. Phys. B, {\bf 680} (2004) 375;
Nucl. Phys. B, {\bf 693} (2004) 134;
G.~Heinrich, Int. J. of Modern Phys. A,  {\bf 23} (2008) 10.
[arXiv:0803.4177].

\bibitem{BognerWeinzierl}
C.~Bogner and S.~Weinzierl,
Comput. Phys. Commun.  {\bf 178} (2008) 596
[arXiv:0709.4092 [hep-ph]];
Nucl. Phys. Proc. Suppl.  {\bf
183} (2008) 256 [arXiv:0806.4307 [hep-ph]].

\bibitem{FIESTA}
  A.~V.~Smirnov and M.~N.~Tentyukov,
  Comput.\ Phys.\ Commun.\  {\bf 180} (2009) 735
  [arXiv:0807.4129 [hep-ph]].

\bibitem{FIESTA2}
A.~V.~Smirnov, V.~A.~Smirnov and M.~Tentyukov,
  arXiv:0912.0158.

\bibitem{MB1}
V.~A.~Smirnov, Phys. Lett.  B  {\bf 460}  (1999) 397;\\
J.~B.~Tausk, Phys. Lett.  B  {\bf 469}  (1999) 225;\\

\bibitem{MB2}
M.~Czakon, Comput.\ Phys.\ Commun.\  {\bf 175} (2006) 559; \\
A.~V.~Smirnov and V.~A.~Smirnov,
JHEP {\bf 05} (2009) 004
[arXiv:0901.0386 [hep-ph]].

\bibitem{books2}
V.~A.~Smirnov, {\em Evaluating Feynman Integrals}, Springer Tracts
Mod.\ Phys.\  {\bf 211} (2004) 1; \\
V.~A.~Smirnov, {\em Feynman integral calculus},
Berlin, Germany: Springer (2006) 283~p.

\bibitem{PSLQ}
 H.R.P.~Ferguson, D.H.~Bailey and S.~Arno,
  Math.\ Comput.\  {\bf 68}, (1999) 351,
  NASA--Ames~Technical Report,
  NAS--96--005.
\end{thebibliography}
\end{document}